\documentclass{article}
\usepackage{sw20lart}

\input{tcilatex}
\begin{document}

\title{Maximum Entropy, Fluctuations and Priors\thanks{%
Presented at MaxEnt 2000, the 20th International Workshop on Bayesian
Inference and Maximum Entropy Methods (July 8-13, 2000, Gif-sur-Yvette,
France).}}
\author{Ariel Caticha \\
{\small Department of Physics, University at Albany-SUNY, }\\
{\small \ Albany, NY 12222, USA.\thanks{%
E-mail: ariel@cnsvax.albany.edu or Ariel.Caticha@albany.edu}}}
\date{}
\maketitle

\begin{abstract}
The method of maximum entropy (ME) is extended to address the following
problem: Once one accepts that the ME distribution is to be preferred over
all others, the question is to what extent are distributions with lower
entropy supposed to be ruled out. Two applications are given. The first is
to the theory of thermodynamic fluctuations. The formulation is exact,
covariant under changes of coordinates, and allows fluctuations of both the
extensive and the conjugate intensive variables. The second application is
to the construction of an objective prior for Bayesian inference. The prior
obtained by following the ME method to its inevitable conclusion turns out
to be a special case ($\alpha =1$) of what are currently known under the
name of entropic priors.
\end{abstract}

\section{Introduction}

The goal of inductive inference is to update a prior probability
distribution to a posterior distribution when new information becomes
available. The problem is to process this information in a systematic and
objective way. When the information is in the form of constraints on the
family of conceivable posterior distributions, there is one inference
procedure that is singled out by requirements of universality, objectivity,
and consistency: it is the method of maximum entropy (ME) \cite{Jaynes57}.
The standard justification relies on interpreting entropy, through the
Shannon axioms, as a measure of the amount of uncertainty in a probability
distribution \cite{Shannon48}, but this justification is not entirely
unobjectionable. A relatively minor problem is that the Shannon axioms refer
to discrete probability distributions rather than continuous ones. A more
serious one is that it is not clear that they provide the only way to define
the notion of uncertainty. This has motivated a number of attempts to
justify the ME method directly, without invoking questionable measures of
uncertainty \cite{ShoreJohnson80}\cite{Skilling88}. They have established
the value of the concept of entropy irrespective of any interpretation in
terms of heat, or disorder, or uncertainty. In these approaches entropy is
purely a tool for consistent reasoning; strictly, \emph{entropy needs no
interpretation}. A welcome by-product has been that the entropy thus defined
turns out to be directly applicable to continuous distributions. In Sect. 2,
as background for the subject of this paper, we present a brief outline of
one such `no-interpretation' approach. Except for one slight modification we
follow Ref.\cite{Skilling88} closely.

The main body of the paper addresses three problems. The unifying element is
the particular form of the constraints; unlike most applications of the ME
method the constraints are not in the form of known expectation values of
certain variables.

The first problem we tackle provides an extension of the ME method itself.
Once one accepts that the maximum entropy distribution is to be preferred
over all others, the question is to what extent, how strongly, are
distributions with lower entropy supposed to be ruled out. In statistical
mechanics the answer to this question is well known. It was first obtained
on the basis of combinatorial arguments in the pioneering work of Boltzmann,
it is the foundation on which Einstein formulated his theory of fluctuations
and Onsager erected his theory of irreversible processes. More recently it
was explored by Jaynes \cite{Jaynes79}. Our goal is to show (Sect. 3) that
the answer can be obtained entirely from within the ME framework, without
appeals to combinatorics, to large systems, or other forms of intuitive
and/or approximate arguments.

The second problem turns out to be a special case of the first: we are
concerned with the theory of fluctuations. The starting point for the
standard theory (see \emph{e.g.} Ref.\cite{Landau77}) is Einstein's
inversion of Bolzmann's formula $S=k\log W$ to obtain the probability of a
fluctuation in the form $W\sim \exp S/k$. A careful justification, however,
reveals a number of approximations which, for most purposes, are legitimate
and work very well. Later developments including the method of cumulants,
the renormalization group, and the connection to non-equilibrium
thermodynamics succeeded in clarifying most of the remaining conceptual and
calculational issues.

A re-examination of fluctuation theory from the point of view of ME is,
however, valuable. Our general conclusion (Sect. 4) is that the ME point of
view allows exact formulations; in fact, it is clear that deviations from
the canonical predictions can be expected, although in general they will be
negligible. Other advantages of the ME\ approach include the explicit
covariance under changes of coordinates, the absence of restrictions to the
vicinity of equilibrium or to large systems, and the conceptual ease with
which one deals with fluctuations of both the extensive as well as their
conjugate intensive variables. This last point is an important one: within
the canonical distribution the extensive variables are random variables
while the intensive ones are fixed parameters, they do not fluctuate. There
are, however, several contexts in which it makes sense to talk about
fluctuations of the conjugate variables. We discuss the standard scenario of
an open system that can exchange say, energy, with its environment. An
altogether different interpretation, which we will not discuss here, is to
consider fluctuations in conjugate variables as uncertainties in the
estimation of parameters \cite{Tikochinsky84}.

The third, and last problem we address is that of obtaining an objective
prior for use in Bayes' theorem. The goal is to obtain information about an
unknown quantity $\theta $ on the basis of the observed value of another
quantity $x$ and of a presumably known relation between $x$ and $\theta $.
This is achieved through Bayes' theorem, $p(\theta |x)\propto \pi (\theta
)p(x|\theta )$. The relation between $x$ and $\theta $ is supplied by a
known model $p(x|\theta )$; previous knowledge about $\theta $ is codified
into the prior probability $\pi (\theta )$.

The selection of a definite prior is a famously controversial issue. It has
generated an enormous literature \cite{Kass96}. The difficulty lies not so
much in a lack of knowledge about $\theta $, but rather in that this
knowledge is sometimes vague: it is not clear how to codify it in an
objective way. Faced with this difficulty one reasonable attitude is to
admit subjectivity, and recognize that different individuals may
legitimately translate the same vague information into different prior
distributions.

An alternative attitude has been to seek some objectivity by demanding
properties such as invariance under reparametrization or other symmetry
transformations. Considerable effort has been spent searching for that
special state of knowledge characterized by complete ignorance, and
accordingly, there are a number of proposals based on the notion of missing
information \cite{Jaynes68}-\cite{Rodriguez98}. In the end, it may turn out
that such a search is misguided; non-informative priors might not exist \cite
{Bernardo97}.

A more positive, direct approach is to identify the information that we do
in fact possess and then find an objective way to take it into account.
Remarkably, it turns out that the very conditions that led us to contemplate
using Bayes' theorem constitute information that can be objectively
translated into a prior using the ME method. The prior thus obtained (Sect.
5) turns out to be one particular member of the family of distributions
known as ``entropic priors.'' The name and the first derivation of this
family for the case of discrete distributions are due to Skilling \cite
{Skilling89}. The generalization to the continuous case and further
elaborations by Rodr\'{\i }guez appear in Ref.\cite{Rodriguez89}. The
immediate motivation for the present work is found in Ref.\cite{Rodriguez98}.

\section{The logic behind the ME method}

Let our beliefs about $x\in X$ be codified in a probability distribution $%
m(x)$. When new information becomes available we want to revise $m(x)$ to a
posterior distribution $p(x)$. The ME method is designed to guide us in
selecting $p(x)$ when the new information is in the form of a specification
of the set of acceptable posterior distributions. The \emph{information} is
just a constraint on the region in the space of all distributions where the
search will be carried out. (These constraints can, but need not, be linear.)

The selection is carried out by ranking the probability distributions
according to increasing \emph{preference}. Two desirable features to be
imposed on this ranking scheme are the following. The first is a
transitivity requirement: if distribution $p_1$ is preferred over
distribution $p_2$, and $p_2$ is preferred over $p_3$, then $p_1$ is
preferred over $p_3$. Such transitive rankings can be implemented by
assigning a real number $S[p]$ to each $p(x)$, in such a way that if $p_1$
is preferred over $p_2$, then $S[p_1]>S[p_2]$. The real number $S[p]$ will
be called the entropy of $p(x)$. (Thus entropies are real numbers by
design.) The selected $p$ will be that which maximizes $S[p]$. (Thus \emph{%
maximum} entropy.)

The problem of finding the functional form of $S[p]$ brings us to the second
desirable feature to be imposed on the ranking scheme. We are looking for a
general rule of inference; the ranking procedure, the rule $S[p]$, must be
of universal applicability: the \emph{same} rule must apply to a variety of
different cases. If we happen to know what the selected distribution should
be in a certain special case, then this knowledge can be used to constrain
the form of $S[p]$. If enough special cases are known $S[p]$ will be
completely determined. These special cases -- the so-called axioms -- must,
by their very nature, be self-evident.

Four axioms are listed below. They all reflect the conviction that changing
one's mind is a serious matter, that one should only update those aspects of
one's beliefs for which hard evidence has been supplied.

\textbf{Axiom 1: Subdomain independence}. If the space $X$ is divided into
non-overlapping subdomains $D_i$, and information is given about $p(x)$ for $%
x\in $ $D_1$, the selection procedure should \emph{only} revise the
(relative) values of $p(x)$ for $x\in D_1$. If the evidence makes no
reference to $x\notin D_1$ those values should be left unchanged.
Non-overlapping subdomains are independent. The power of this axiom lies in
that the choice of subdomains $D_i$ is arbitrary, the consequence is that
non-overlapping domains contribute additively to $S[p].$

\textbf{Axiom 2: Coordinate invariance.} The ranking should not depend on
the particular system of coordinates being used. The coordinates used to
label the points $x$ do not carry any information. The consequence of this
axiom is that the expression for $S[p]$ will involve coordinate invariants
such as $dx\,p(x)$ and ratios such as $p(x)/m(x)$, where the function $m(x)$
is, at this point, any arbitrary measure.

\textbf{Axiom 3:\ Subsystem independence}. If a system is composed of two
subsytems, $x=(x_1,x_2)\in X=X_1\times X_2$, the selection procedure should
introduce no correlations for which there was no evidence either in the
measure or in the constraints. As a consequence of this axiom a logarithm
appears in the expression for $S[p]$.

\textbf{Axiom 4: Objectivity}. If there is no new information there is no
reason to change one's mind: when there are no constraints the selected
posterior distribution should coincide with the prior distribution. The
arbitrariness in $m(x)$ is now removed: $m(x)$ is the prior distribution.

The overall consequence of these axioms (for a proof see \cite{Skilling88})
is that probability distributions should be ranked according to their
entropy, 
\begin{equation}
S[p]=-\int dx\,p(x)\log \frac{p(x)}{m(x)}.  \label{S[p]}
\end{equation}

Choosing the prior $m(x)$ can be tricky. When there is no information
leading us to prefer one microstate of a physical system over another we
might as well assign equal prior probability to each state. Thus it is
reasonable to choose the density of states as the prior distribution $m(x)$;
the invariant $m(x)dx$ is the number of microstates in $dx$. This is the
basis for statistical mechanics.

Other examples of relevance to physics arise when there is no reason to
prefer one region of the space $X$ over another. Then we should assign the
same prior probability to regions of the same ``volume,'' and we can choose $%
\int_Rdx\,m(x)$ to be the volume of a region $R$ in the space $X$. On the
basis of this choice of prior one can \emph{derive} a considerable amount of
the formalism of quantum mechanics. This includes the ``postulates'' that
quantum states form a Hilbert space, that probabilities are computed through
the Born rule, and that time evolution is unitary \cite{Caticha99}.

Notice that through the measure $m(x)$ Laplace's principle of insufficient
reason still plays a role, albeit in a somewhat modified form. Thus,
subjectivity has not been eliminated. Just as with Bayes' theorem, what is
objective here is the manner in which information is processed, not the
initial probability assignments.

\section{Extending the ME method}

Let $X$ be the space of microstates $x$ of a physical system ($x\in X$), and
let $m(x)dx$ be the number of microstates in the range $dx$. (Although in
this and in the next section we tend to drift into the language of
statistical mechanics it will be clear that the central idea is easily
exported to other contexts.) We assume that the expected values $A^{{}\alpha
}$ of some $n_A$ variables $a^{{}\alpha }(x)$ ($\alpha =1,2,\ldots ,n_A$)
are known, 
\begin{equation}
\left\langle a^{{}\alpha }\right\rangle =\int dx\,p(x)a^{{}\alpha
}(x)=A^{{}\alpha }\,.  \label{Aalpha}
\end{equation}
This limited information will certainly not be sufficient to answering all
questions that one could conceivably ask about the system. Therefore, we
make the further assumption that the set $\{a^{{}\alpha }\}$ has not been
randomly chosen, that it has been carefully selected because previous
experience indicates the information in (\ref{Aalpha}) is relevant for our
purposes.

The probability distribution $p_0(x)$ that best reflects the prior
information contained in $m(x)$ updated by the information $A^{{}\alpha }$
is obtained by maximizing (\ref{S[p]}) subject to the constraints (\ref
{Aalpha}). The result is 
\begin{equation}
p_0(x)=\frac 1Z\,m(x)\,e^{-\lambda _{{}\alpha }a^{{}\alpha }(x)},
\label{pzero}
\end{equation}
where the partition function $Z$ and the Lagrange multipliers $\lambda
_{{}\alpha }$ are given by 
\begin{equation}
Z(\lambda )=\int dx\,m(x)\,e^{-\lambda _{{}\alpha }a^{{}\alpha }(x)}\quad 
\text{and}\quad -\frac{\partial \log Z}{\partial \lambda _{{}\alpha }}%
=A^{{}\alpha }\,.  \label{Z and lambda}
\end{equation}

The question we address concerns the extent to which the maximum entropy
distribution $p_0(x)$ should be preferred over other distributions with
lower entropy. Consider a family of distributions $p(x|\theta )$ which
depends on a finite, though arbitrarily large, number $n_{{}\theta }$ of
parameters $\theta ^i$ ($i=1,2,\ldots ,n_{{}\theta }$) and which includes $%
p_0(x)$ as one of its elements. We can choose the parameters $\theta ^i$ so
that $p(x|\theta =0)=p_0(x)$.

The question about the extent that $p(x|\theta =0)$ is to be preferred over $%
p(x|\theta \neq 0)$ can be phrased more suggestively as follows. To \emph{%
what extent do we believe} that the correct selection should be $\theta =0$
rather than $\theta \neq 0$? Thus, our question has metamorphosed into an
inquiry about a degree of belief: the probability of $\theta $, $\pi (\theta
)$. In fact, we should go further back. The original endeavor which led us
to use the ME method in the first place was a question about the probability
of $x$, now we are actually asking about the probability of ``$x$ and $%
\theta $.'' We want not just $p(x)$ but rather $p(x,\theta )$; asking about
the reliability of the answer $p_0(x)$ has led us to expand the universe of
discourse from $X$ to $X\times \Theta $ where $\Theta $ is the space of
parameters $\theta $. It is remarkable that this is precisely the kind of
question the ME method is designed to answer; the strategy is to determine
the distribution $p(x,\theta )$ by maximizing an entropy subject to whatever
constraints are known to hold. To proceed we must address two questions:
precisely what is the form of the entropy to be maximized, and second, what
are the constraints on $p(x,\theta )$.

No definition of entropy is complete until a measure over the space in
question ($X\times \Theta $) is specified. Our starting point is that a
priori there is no known connection between $x$ and the arbitrary set of
parameters $\theta $. Since a measure must not by itself introduce
correlations for which there is no evidence, the prior measure $m(x,\theta )$
must be a product, $m(x)\mu (\theta )$ of the known density of states $m(x)$
and a still unknown measure $\mu (\theta )$ over the space $\Theta $. Thus,
the entropy to be maximized is 
\begin{equation}
\sigma [p]=-\int dx\,d\theta \,p(x,\theta )\,\log \frac{p(x,\theta )}{%
m(x)\mu (\theta )},
\end{equation}

Next we incorporate the crucial piece of information that gives meaning to
the parameters $\theta $ and establishes the relation between $\theta $ and $%
x$: the conditional probability $p(x|\theta )$ is known. This has two
consequences: First, the joint distribution $p(x,\theta )$ is constrained to
be of the form $\pi (\theta )p(x|\theta )$. Notice that this constraint is
not in the usual form of an expectation value. Second, the ambiguity in the
choice of the measure $\mu (\theta )$ in $\Theta $ is resolved. The family
of distributions $p(x|\theta )$ induces a natural distance in the space $%
\Theta $: $d\ell ^2=g_{ij}d\theta ^id\theta ^j$, where $g_{ij}$ is the
Fisher-Rao metric \cite{Amari85}\cite{Caticha00}, 
\begin{equation}
g_{ij}=\int dx\,p(x|\theta )\frac{\partial \log p(x|\theta )}{\partial
\theta ^i}\frac{\partial \log p(x|\theta )}{\partial \theta ^j}.
\label{Fisher metric}
\end{equation}
Accordingly we choose $\mu (\theta )=g^{1/2}(\theta )$, where $g(\theta )$
is the determinant of $g_{ij}$. Having identified the measure and the
constraints, we allow the ME method to take over.

The preferred distribution $p(x,\theta )$ is chosen by varying $\pi (\theta
) $ to maximize 
\begin{equation}
\sigma [\pi ]=-\int dx\,d\theta \,\pi (\theta )p(x|\theta )\,\log \frac{\pi
(\theta )p(x|\theta )}{g^{1/2}(\theta )m(x)}.  \label{sigma[pi]}
\end{equation}
Assume $\,p(x|\theta )$ is normalized, $\int dx\,p(x|\theta )=1$. Maximizing
(\ref{sigma[pi]}) with respect to variations $\delta \pi (\theta )$ such
that $\int d\theta \,\pi (\theta )=1$, yields 
\begin{equation}
0=\int \,d\theta \left( -\log \frac{\pi (\theta )}{g^{1/2}(\theta )}%
+S(\theta )+\log \zeta \right) \,\delta \pi (\theta )\,,
\end{equation}
where the required Lagrange multiplier has been written as $1-\log \zeta $,
and 
\begin{equation}
S(\theta )=-\int \,dx\,p(x|\theta )\log \frac{p(x|\theta )}{m(x)}.
\label{Stheta}
\end{equation}
Therefore the probability that the value of $\theta $ should lie within the
small volume $g^{1/2}(\theta )d\theta $ is 
\begin{equation}
\pi (\theta )d\theta =\frac 1\zeta \,\,e^{S(\theta )}g^{1/2}(\theta )d\theta
\quad \text{with\quad }\zeta =\int d\theta \,g^{1/2}(\theta )\,e^{S(\theta
)}.  \label{main}
\end{equation}
Equation (\ref{main}) is our main result. It tells us that, as expected, the
preferred value of $\theta $ is that which maximizes the entropy $S(\theta )$
because this maximizes the scalar probability density $\exp S(\theta )$. But
it also tells us the degree to which values of $\theta $ away from the
maximum are ruled out. For macroscopic systems the preference for the ME
distribution can be overwhelming.

Note that the density $\exp S(\theta )$ is a scalar function and the
presence of the Jacobian factor $g^{1/2}(\theta )$ makes Eq.(\ref{main})
manifestly invariant under changes of the coordinates $\theta ^i$ in the
space $\Theta $.

\section{Fluctuations}

Fluctuations of the variables $a^{{}\alpha }(x)$ or of any function $b(x)$
of the microstate $x$ are usually computed in terms of the various moments
of the canonical ME distribution $p_0(x)$ given by Eqs.(\ref{pzero}-\ref{Z
and lambda}) (see, however, Ref.\cite{Jaynes78}). Within this context all
expected values, such as the constraints $\left\langle a^{{}\alpha
}\right\rangle =$ $A^{{}\alpha }$ and the entropy $S(A)$ itself are fixed,
they do not fluctuate. The corresponding conjugate variables, the Lagrange
multipliers $\lambda _{{}\alpha }=\partial S/\partial A^{{}\alpha }$, do not
fluctuate either.

The standard way to make sense of $\lambda $ fluctuations is to couple the
system of interest to a second system, a bath, and allow exchanges of the
quantities $a^{{}\alpha }$. All quantities referring to the bath will be
denoted by primes: microstates $x^{\prime }$, density of states $m^{\prime
}(x^{\prime })$, variables $a^{\prime \alpha }(x^{\prime })$, etc. Even
though the overall expected value $\left\langle a^{{}\alpha }+a^{\prime
\alpha }\right\rangle =A_T^{{}\alpha }$ of the combined system plus bath is
fixed, the individual expected values $\left\langle a^{{}\alpha
}\right\rangle =$ $A^{{}\alpha }$ and $\left\langle a^{\prime \alpha
}\right\rangle =$ $A^{\prime \alpha }=A_T^{{}\alpha }-A^{{}\alpha }$ are
allowed to fluctuate. The ME distribution $p_0(x,x^{\prime })$ that best
reflects the prior information contained in $m(x)$ and $m^{\prime
}(x^{\prime })$ updated by information on the total $A_T^{{}\alpha }$ is 
\begin{equation}
p_0(x,x^{\prime })=\frac 1{Z_0}\,m(x)m^{\prime }(x^{\prime })\,e^{-\lambda
_{0\alpha }\left( a^{{}\alpha }(x)+a^{\prime \alpha }(x^{\prime })\right) }.
\end{equation}
But less than ME distributions are not totally ruled out; to explore the
possibility that the quantity $A_T$ is distributed between the two systems
in a less than optimal way we consider distributions $p(x,x^{\prime },A)$
constrained to the form 
\begin{equation}
p(x,x^{\prime },A)=\pi (A)p(x|A)p(x^{\prime }|A_T-A),  \label{pi.pp}
\end{equation}
where 
\begin{equation}
p(x|A)=\frac 1{Z(\lambda )}\,m(x)\,e^{-\lambda _{{}\alpha }a^{{}\alpha }(x)}.
\end{equation}
The corresponding entropy is 
\begin{equation}
S(A)=\log Z(\lambda )+\lambda _{{}\alpha }A^{{}\alpha }\,,
\end{equation}
with $\lambda _{{}\alpha }$ and $Z(\lambda )$ given by Eq.(\ref{Z and lambda}%
). Analogous expressions hold for the primed quantities. The formalism
simplifies considerably when the bath is large enough that exchanges of $A$
do not affect it, and $\lambda ^{\prime }$ remains fixed at $\lambda _0$.
Then 
\begin{equation}
S^{\prime }(A_T-A)=\log Z^{\prime }(\lambda _0)+\lambda _{0\alpha }\left(
A_T^{{}\alpha }-A^{{}\alpha }\right) =\func{const}-\lambda _{0\alpha
}A^{{}\alpha }\text{.}
\end{equation}

The probability that the value of $A$ fluctuates into a small volume $%
g^{1/2}(A)dA$ is given by our main result Eq.(\ref{main}), 
\begin{equation}
\pi (A)dA=\frac 1{\zeta (\lambda _0)}\,\,e^{S(A)-\lambda _{0\alpha
}A^{{}\alpha }}g^{1/2}(A)dA\quad \text{where}\quad g_{\alpha \beta }=-\frac{%
\partial ^2S(A)}{\partial A^{{}\alpha }\partial A^{{}\beta }}\,,
\label{fluctuations}
\end{equation}
and $\zeta (\lambda _0)$ is a suitably defined normalization. To the extent
that the right choice of variables has been made, Eq.(\ref{fluctuations}) is
exact.

An important difference with the usual theory stems from the presence of the
Jacobian factor $g^{1/2}(A)$. This is required by coordinate invariance and
can lead to small deviations from the canonical predictions. The quantities $%
\left\langle \lambda _{{}\alpha }\right\rangle $ and $\left\langle
A^{{}\alpha }\right\rangle $ may be close but will not in general coincide
with the quantities $\lambda _{0\alpha }$ and $A_0^{{}\alpha }$ at the point
where the scalar probability density attains its maximum. When this maximum
is very sharp and in its vicinity the Jacobian can be considered constant
the usual results \cite{Landau77} follow. The remaining difficulties are
purely computational and of the kind that can in general be tackled
systematically using the method of steepest descent to evaluate the
appropriate generating function.

Since we are not interested in variables referring to the bath we can
integrate Eq.(\ref{pi.pp}) over $x^{\prime }$, and use the distribution $%
p(x,A)=\pi (A)p(x|A)$ to compute various moments. As an example, the
correlation between $\delta \lambda _{{}\alpha }=\lambda _{{}\alpha
}-\left\langle \lambda _{{}\alpha }\right\rangle $ and $\delta a^{{}\beta
}=a^{{}\beta }-\left\langle A^{{}\beta }\right\rangle $ or $\delta
A^{{}\beta }=A^{{}\beta }-\left\langle A^{{}\beta }\right\rangle $ is 
\begin{equation}
\left\langle \delta \lambda _{{}\alpha }\delta a^{{}\beta }\right\rangle
=\left\langle \delta \lambda _{{}\alpha }\delta A^{{}\beta }\right\rangle =-%
\frac{\partial \left\langle \lambda _{{}\alpha }\right\rangle }{\partial
\lambda _{0\beta }}+\left( \lambda _{0\alpha }-\left\langle \lambda
_{{}\alpha }\right\rangle \right) \left( A_0^{{}\beta }-\left\langle
A^{{}\beta }\right\rangle \right) .
\end{equation}
When the differences $\lambda _{0\alpha }-\left\langle \lambda _{{}\alpha
}\right\rangle $ or $A_0^{{}\beta }-\left\langle A^{{}\beta }\right\rangle $
are negligible one obtains the usual expression, $\left\langle \delta
\lambda _{{}\alpha }\delta a^{{}\beta }\right\rangle \approx -\delta
_{{}\alpha }^{{}\beta }\,$.

\section{Entropic priors}

The last problem we address is that of obtaining a prior $\pi (\theta )$ for
use in Bayes' theorem, $p(\theta |x)\propto \pi (\theta )p(x|\theta )$. The
traditional approach has been to attempt to determine or at least to
constrain $\pi (\theta )$ by requiring that it be non-informative, that it
satisfy coordinate invariance, and so on. The seemingly innocuous but
fruitful new idea proposed by Rodr\'{\i }guez \cite{Rodriguez98} is to focus
attention on $p(x,\theta )$ instead of $\pi (\theta )$. One could well
wonder whether this makes any difference. After all, once $p(x|\theta )$ is
known, $p(x,\theta )$ can be calculated from $\pi (\theta )$ and vice versa.

It makes a huge difference. The selection of a preferred distribution using
the ME method demands that one specify in which space the search will be
conducted. Being a consequence of the product rule, Bayes' theorem requires
that $p(x,\theta )$ be defined and that assertions such as ``$x$ \emph{and} $%
\theta $'' be meaningful. The relevant universe of discourse is neither $X$
nor $\Theta $, but the product $X\times \Theta $.

The complete specification of the space $X\times \Theta $ requires a measure 
$m(x,\theta )$. At this point we do not know anything about the variables $%
\theta $, they are totally arbitrary. To the extent that no relation between 
$x$ and $\theta $ is known, the measure must be the product $m(x)\mu (\theta
)$ of the separate measures in the spaces $X$ and $\Theta $. Indeed, the
distribution that maximizes 
\begin{equation}
\sigma [p]=-\int dx\,d\theta \,p(x,\theta )\,\log \frac{p(x,\theta )}{%
m(x)\mu (\theta )},
\end{equation}
is $p(x,\theta )\propto m(x)\mu (\theta )$; it is such that data about $x$
tells us nothing about $\theta $. In what follows we assume that $m(x)$ is
known; \emph{this is part of understanding what data it is that has been
collected}. The measure $\mu (\theta )$ remains undetermined.

Next we incorporate the crucial piece of information: in order to infer
something about $\theta $ on the basis of a measurement of $x$, a relation
between $x$ and $\theta $ must exist. The relation is supplied by the model $%
p(x|\theta )$. This constrains the joint distribution $p(x,\theta )$ to be
of the form $\pi (\theta )p(x|\theta )$ and removes the ambiguity in the
choice of $\mu (\theta )$. As mentioned before, there is a natural choice $%
\mu (\theta )=g^{1/2}(\theta )$, where $g(\theta )$ is the determinant of
the Fisher-Rao metric $g_{ij}$. Having identified the space, the measure,
and the constraints, the ME method gives the probability $\pi (\theta )$
that the value of $\theta $ should lie within the small volume $%
g^{1/2}(\theta )d\theta $. It is our previous main result, Eq.(\ref{main}), 
\begin{equation}
\pi (\theta )d\theta \propto \,\,e^{S(\theta )}g^{1/2}(\theta )d\theta .
\label{main2}
\end{equation}

It is remarkable that the ingredients that have been used are precisely
those that led us to consider using Bayes' theorem in the first place. Once
the model is known, which means that the data space $X$, its measure $m(x)$,
and the conditional distribution $p(x|\theta )$ are given, the prior
probability $\pi (\theta )$ is unambiguously determined.

We emphasize that $\pi (\theta )$ in Eq.(\ref{main2}) is not the least
informative distribution, it is the distribution \emph{after} we learn $%
p(x|\theta )$. The distribution \emph{before} we learn $p(x|\theta )$ is $%
\mu (\theta )$. We do not know it; this is truly noninformative.

No doubt the reader recognizes that essentially the same argument has been
given twice, first in Sect. 3 and then here. There is a reason for this
repetition. It was not a priori obvious (at least to this author) that there
could have existed a relation between, say, the theory of thermodynamic
fluctuations and the problem of selecting priors in Bayesian inference. They
are most definitely not the same problem; the meanings of the various
symbols and the motivations driving our interests in these questions do not
coincide. It was therefore not at all clear that exactly the same
mathematical formalism could provide the solution to both. Two verbal
justifications, rather than just one, were needed.

The prior $\pi (\theta )$ in Eq.(\ref{main2}) is a member of the family of
distributions labelled by the real parameter $\alpha $, 
\begin{equation}
\pi (\theta ,\alpha )=\frac 1{\zeta (\alpha )}\,\,e^{\alpha S(\theta
)}g^{1/2}(\theta ),\quad \text{with\quad }\zeta (\alpha )=\int d\theta
\,g^{1/2}(\theta )\,e^{\alpha S(\theta )},  \label{pialpha}
\end{equation}
which are known as entropic priors \cite{Skilling89}-\cite{Rodriguez98}. The
ME approach has unambiguously selected the $\alpha =1$ member. Indeed, it is
easy to check that values $\alpha \neq 1$ do not maximize the $\sigma $
entropy, $\sigma [\pi (\theta ,1+\varepsilon )]\leq \sigma [\pi (\theta ,1)]$%
.

The $\alpha =1$ entropic prior has, in the past, led to manifestly
reasonable results. Examples include the entropic prior for the family of
Gaussians \cite{Rodriguez98}, and the distribution dual to the
Maxwell-Boltzmann distribution \cite{Tikochinsky84}. The justifications
given for these two cases are totally independent of each other and of ours;
both are instances of Jeffreys' prior for scale parameters \cite{Kass96}. On
the other hand, values $\alpha \neq 1$ have also been used. To investigate
this further we consider experiments that can be repeated.

Experiments need not be repeatable. Assume, however, that successive
repetitions are possible and that they happen to be independent. Suppose, to
be specific, that the experiment is performed twice so that the space of
data $X\times X=X^2$ consists of the possible outcomes $x_1$ and $x_2$.
Suppose further that $\theta $ is not a ``random'' variable; the value of $%
\theta $ is fixed but unknown. Then the joint distribution in the space $%
X^2\times \Theta $ is 
\begin{equation}
p(x_1,x_2,\theta )=\pi ^{(2)}(\theta )\,p(x_1,x_2|\theta )=\pi ^{(2)}(\theta
)p(x_1|\theta )p(x_2|\theta ),
\end{equation}
and the appropriate $\sigma $ entropy is 
\begin{equation}
\sigma ^{(2)}[\pi ]=-\int dx_1\,dx_2\,d\theta \,p(x_1,x_2,\theta )\,\log 
\frac{p(x_1,x_2,\theta )}{\sqrt{g^{(2)}(\theta )}\,m(x_1)m(x_2)},
\end{equation}
where $g^{(2)}(\theta )$ is the determinant of the Fisher-Rao metric for $%
p(x_1,x_2|\theta )$. From Eq.(\ref{Fisher metric}) it follows that $%
g_{ij}^{(2)}=2g_{ij}$ so that $g^{(2)}(\theta )=2^dg(\theta )$, $d$ being
the dimension of $\theta $. Maximizing $\sigma ^{(2)}[\pi ]$ subject to $%
\int \,d\theta \,\pi ^{(2)}(\theta )=1$ we get 
\begin{equation}
\pi ^{(2)}(\theta )=\frac 1{Z^{(2)}}\,g^{1/2}(\theta )\,e^{S^{(2)}(\theta
)}=\frac 1{Z^{(2)}}\,g^{1/2}(\theta )\,e^{2S^{(1)}(\theta )},
\end{equation}
where $S^{(2)}(\theta )=2S^{(1)}(\theta )\equiv 2S(\theta )$ is the entropy
of $\,p(x_1,x_2|\theta )$. The generalization to $n$ repetitions of the
experiment, with data space $X^n$, is immediate: the ME prior $\pi
^{(n)}(\theta )$ is obtained replacing $S^{(2)}(\theta )$ with $%
S^{(n)}(\theta )=nS^{(1)}(\theta )$. The coefficient in front of $%
S^{(n)}(\theta )$ remains $\alpha =1$ and the prior $\pi ^{(n)}(\theta )$
differs from $\pi ^{(1)}(\theta )$. This is puzzling. Do we have to revise
our prior as more data comes in? In fact, for large $n$ the prior $\pi
^{(n)}(\theta )$ above becomes manifestly wrong: the exponential preference
for the value of $\theta $ that maximizes $S^{(1)}(\theta )$ becomes so
pronounced that no amount of data to the contrary can successfully overcome
its effect.

Repeatable experiments present us with a problem, but how do we deny
preferred status to $\alpha =1$ without simultaneously challenging the ME
principle itself? There is one way out of this dilemma. Readers of Jaynes'
work will surely recognize the following argument: we have been conducting
our search with the wrong constraint. There is something we know about
repeatable experiments that we have not incorporated into the ME procedure
above. I propose it is this: when we say an experiment can be repeated say,
twice, $n=2$, we actually know more than just $p(x_1,x_2|\theta
)=\,p(x_1|\theta )p(x_2|\theta )$. We also know that forgetting or
discarding the value of say $x_2$, yields an experiment that is totally
indistinguishable from the single, $n=1$, experiment. This \emph{additional}
information is quantitatively expressed by the constraint $\int
dx_2\,p(x_1,x_2,\theta )=p(x_1,\theta )$, or equivalently 
\begin{equation}
\int dx_2\,\pi ^{(2)}(\theta )p(x_1|\theta )p(x_2|\theta )=\pi ^{(1)}(\theta
)p(x_1|\theta )\,,
\end{equation}
which leads to $\pi ^{(2)}(\theta )=\pi ^{(1)}(\theta )$. In the general
case we get the manifestly reasonable result $\pi ^{(n)}(\theta )=\pi
^{(n-1)}(\theta )=\ldots =\pi ^{(1)}(\theta )$; the undesired dependence on $%
n$ has been eliminated.

The conclusion is that our result Eq.(\ref{main2}) stands: $\alpha =1$ is
the default value. Unless there is positive evidence to the contrary, the
entropic prior with $\alpha =1$ should be preferred. But, of course, the
results of Sect. 3 apply here too. The preference for maximum entropy is not
absolute: $\alpha =1$ is just the maximum $\sigma $ distribution, and values
of $\alpha $ corresponding to less than maximum $\sigma $ are not totally
ruled out.

\section{Final remarks}

The method of maximum entropy has been extended to give a quantitative
determination of the degree to which distributions with lower entropy are
ruled out. The same idea was used to extend the theory of thermodynamic
fluctuations and in the construction of priors for Bayesian inference. That
a connection between these two historically independent topics should at all
exist is in itself quite remarkable.

We conclude with a comment on the reliability of using entropy as a tool for
reasoning. There are several reasons why the ME method could lead to an
absurd answer. One possibility is that there is relevant prior information
that remains unidentified. Another possible reason for failure is a wrong
choice of variables. Choosing the right variables is perhaps the most
serious difficulty in statistical mechanics; in fact, it takes many years of
indoctrination before it is obvious that the Cooper pair wave function is
the right variable to describe superconductivity.

These two possibilities, failure to identify the correct constraints or to
identify the correct variables, \emph{do not reflect} \emph{a flaw of the ME
method itself}. Of course, it is conceivable, that it is the ME axioms that
fail, or that real numbers are not the right way to measure entropy, or even
worse, that there is no universal set of rules for processing information.
But one need not be overly cautious in this last respect. It is clear that
the ME method is applicable to a vast range of problems, and at this point,
there are absolutely no signs that the exploration of this territory is
anywhere near completion.

\noindent \textbf{Acknowledgments- }I am indebted to C. C. Rodr\'{\i}guez
and D. A. Davis for very valuable discussions.


\begin{thebibliography}{99}
\bibitem{Jaynes57}  E. T. Jaynes, ``Information Theory and Statistical
Mechanics'' Phys. Rev. \textbf{106}, 620 and \textbf{108}, 171 (1957).

\bibitem{Shannon48}  C. E. Shannon, Bell Systems Tech. Jour. \textbf{27},
379, 623 (1948); C. E. Shannon and W. Weaver, \emph{The Mathematical Theory
of Communication} (Univ. of Illinois Press, Urbana, 1949); N. Wiener, \emph{%
Cybernetics} (MIT Press, Cambridge, 1948); L. Brillouin, \emph{Science and
Information Theory}, (Academic Press, New York, 1956); S. Kullback, \emph{%
Information Theory and Statistics} (Wiley, New York, 1959).

\bibitem{ShoreJohnson80}  J. E. Shore and R. W. Johnson, ``Axiomatic
derivation of the Principle of Maximum Entropy and the Principle of Minimum
Cross-Entropy,'' IEEE Trans. Inf. Theory \textbf{IT-26}, 26 (1980); Y.
Tikochinsky, N. Z. Tishby and R. D. Levine, Phys. Rev. Lett. \textbf{52},
1357 (1984) and Phys. Rev. \textbf{A30}, 2638 (1984).

\bibitem{Skilling88}  J. Skilling, ``The Axioms of Maximum Entropy'' in 
\emph{Maximum-Entropy and Bayesian Methods in Science and Engineering}, G.
J. Erickson and C. R. Smith (eds.) (Kluwer, Dordrecht, 1988).

\bibitem{Jaynes79}  E. T. Jaynes, ``Concentration of Distributions at
Entropy Maxima,'' reprinted in R. D. Rosenkrantz (ed.), \emph{E. T. Jaynes:
Papers on Probability, Statistics and Statistical Physics} (Reidel,
Dordrecht, 1983).

\bibitem{Landau77}  L. D. Landau and E. M. Lifshitz, \emph{Statistical
Physics} (Pergamon, New York, 1977); H. B. Callen, \emph{Thermodynamics and
an Introduction to Thermostatistics} (Wiley, New York, 1985).

\bibitem{Tikochinsky84}  Y. Tikochinsky and R. D. Levine, J. Math. Phys. 
\textbf{25}, 2160 (1984); G. D. J. Phillies, Am. J. Phys. \textbf{52}, 629
(1984); H. B. Prosper, Am. J. Phys. \textbf{61}, 54 (1993); F. Schl\"{o}gl,
Z. Physik \textbf{244}, 199 (1971) and J. Phys. Chem. Solids \textbf{49},
679 (1988).

\bibitem{Amari85}  S. Amari, \emph{Differential-Geometrical Methods in
Statistics} (Springer-Verlag, 1985).

\bibitem{Caticha00}  For a brief derivation see A. Caticha, ``Change, Time
and Information Geometry,'' in these proceedings.

\bibitem{Jaynes78}  Experimentally measured fluctuations and quadratic
moments need not coincide; the conditions under which they do are discussed
in E. T. Jaynes, ``Where do we stand on maximum entropy?'' in \emph{The
Maximum Entropy Formalism}, ed. by R. D.\ Levine and M. Tribus (MIT Press,
Cambridge, 1978).

\bibitem{Kass96}  For a review with annotated bibliography see \emph{e.g.},
R. E. Kass and L. Wasserman, J. Am. Stat. Assoc. \textbf{91}, 1343 (1996).

\bibitem{Jaynes68}  E. T. Jaynes, IEEE Trans. Syst. Sci. Cybern. Vol. 
\textbf{SSC-4}, 227 (1968); J. M. Bernardo, J. Roy. Stat. Soc. \textbf{41},
113 (1979); A. Zellner, ``Bayesian methods and entropy in economics and
econometrics'' in \emph{Maximum Entropy and Bayesian Methods}, edited by W.
T. Grandy Jr. and L. H. Schick (Kluwer, Dordrecht, 1991).

\bibitem{Skilling89}  J. Skilling, ``Classic Maximum Entropy'' in \emph{%
Maximum Entropy and Bayesian Methods}, J. Skilling (ed.) (Kluwer, Dordrecht,
1989); ``Quantified Maximum Entropy'' in \emph{Maximum Entropy and Bayesian
Methods}, P. F. Foug\`{e}re (ed.) (Kluwer, Dordrecht, 1990).

\bibitem{Rodriguez89}  C. C. Rodr\'{\i }guez, ``The metrics generated by the
Kullback number'' in \emph{Maximum Entropy and Bayesian Methods}, J.
Skilling (ed.) (Kluwer, Dordrecht, 1989); ``Objective Bayesianism and
geometry'' in \emph{Maximum Entropy and Bayesian Methods}, P. F. Foug\`{e}re
(ed.) (Kluwer, Dordrecht, 1990); ``Entropic priors'' in \emph{Maximum
Entropy and Bayesian Methods}, edited by W. T. Grandy Jr. and L. H. Schick
(Kluwer, Dordrecht, 1991); ``Bayesian robustness: a new look from geometry''
in \emph{Maximum Entropy and Bayesian Methods}, G. R. Heidbreder (ed.)
(Kluwer, Dordrecht, 1996).

\bibitem{Rodriguez98}  C. C. Rodr\'{\i }guez, see section 3 of ``Are we
cruising a hypothesis space?'' in \emph{Maximum Entropy and Bayesian Methods}%
, edited by W. von der Linden, V. Dose, R. Fischer and R. Preuss (Kluwer,
Dordrecht, 1999).

\bibitem{Bernardo97}  J. M. Bernardo, T. Z. Irony, N. D. Singpurwalla, J.
Stat. Plan. Inf. \textbf{65}, 159 (1997).

\bibitem{Caticha99}  Ariel Caticha, ``Probability and entropy in quantum
theory,'' in \emph{Maximum Entropy and Bayesian Methods}, ed. by W. von der
Linden et al. (Kluwer, Dordrecht, 1999) (online at
http://xxx.lanl.gov/abs/quant-ph/9808023); ``Insufficient reason and entropy
in quantum theory,'' to appear in Found. Phys. (2000) (online at
http://xxx.lanl.gov/abs/quant-ph/9810074).
\end{thebibliography}
\end{document}